ORIGINAL PAPER

# Exploring the cocrystallization potential of urea and benzamide

Piotr Cysewski[1] · Maciej Przybyłek[1] · Dorota Ziółkowska[2] · Karina Mroczyńska[2]

**Abstract** The cocrystallization landscape of benzamide and urea interacting with aliphatic and aromatic carboxylic acids was studied both experimentally and theoretically. Ten new cocrystals of benzamide were synthesized using an oriented samples approach via a fast dropped evaporation technique. Information about types of known bi-component cocrystals augmented with knowledge of simple binary eutectic mixtures was used for the analysis of virtual screening efficiency among 514 potential pairs involving aromatic carboxylic acids interacting with urea or benzamide. Quantification of intermolecular interaction was achieved by estimating the excess thermodynamic functions of binary liquid mixtures under supercooled conditions within a COSMO-RS framework. The smoothed histograms suggest that slightly more potential pairs of benzamide are characterized in the attractive region compared to urea. Finally, it is emphasized that prediction of cocrystals of urea is fairly direct, while it remains ambiguous for benzamide paired with carboxylic acids. The two known simple eutectics of urea are found within the first two quartiles defined by excess thermodynamic functions, and all known cocrystals are outside of this range belonging to the third or fourth quartile. On the contrary, such a simple separation of positive and negative cases of benzamide miscibility in the solid state is not observed. The difference in properties between urea and benzamide R2,2(8) heterosynthons is also documented by alterations of substituent effects. Intermolecular interactions of urea with para substituted benzoic acid analogues are stronger compared to those of benzamide. Also, the amount of charge transfer from amide to aromatic carboxylic acid and vice versa is more pronounced for urea. However, in both cases, the greater the electron withdrawing character of the substituent, the higher the binding energy, and the stronger the supermolecule polarization via the charge transfer mechanism.

✉ Piotr Cysewski
  piotr.cysewski@cm.umk.pl

[1] Department of Physical Chemistry, Faculty of Pharmacy, Collegium Medicum of Bydgoszcz, Nicolaus Copernicus University in Toruń, Kurpińskiego 5, 85-950 Bydgoszcz, Poland
[2] Research Laboratory, Faculty of Chemical Technology and Engineering, University of Technology and Life Sciences in Bydgoszcz, Seminaryjna 3, 85-326 Bydgoszcz, Poland

## Introduction

Cocrystals represent an interesting and practically important subgroup of multicomponent solids. By definition, they are homogenous crystalline systems containing stoichiometric amounts of one or more neutral molecular species, which are in the solid state under ambient temperature and pressure conditions [1]. Consequently, species such as clathrates, solvates and salts are not usually classified as cocrystals. It is worth mentioning that cocrystallization has been used widely in many fields, including the agrochemistry, electronics, textile and pharmaceutical industries. The latter probably accounts for their the most valuable applications [2]. Supramolecular synthesis has been applied mainly to overcome the poor

Springer

solubility of active pharmaceutical ingredients (API) [3], and consequently improve their bioavailability, mechanical properties [4], stability [5] and other physicochemical properties.

There are several efficient methods of cocrystal preparation, such as solvent-assisted grinding [6–8], anti-solvent crystallization [9–11], slurry cocrystallization [12–14] and solvent evaporation [15, 16] approaches. Generally speaking, cocrystal preparation methods can be divided into two categories, namely kinetic (fast crystal growth) and thermodynamic (slow crystallization) [17–19]. Notably, grinding is the oldest method of cocrystal preparation, and was used for synthesis of the very first molecular complex formed by hydroquinone and quinone [20]. Particularly interesting are methods like spray drying [21–23] and droplet evaporative crystallization (DEC) [24] that involve evaporation of small amounts of solution. They are efficient, fast, environmentally friendly and cost preserving, as documented in our previous studies [24–26]. The formation of oriented crystallite samples on glass surfaces can be analyzed successfully with the aid of instrumental methods such as powder X-ray diffraction (PXRD) and attenuated total reflectance Fourier transform infrared spectroscopy (ATR-FTIR). The main advantage of PXRD measurements of oriented samples is the enhancement of the intensity of diffraction signals coming from the cocrystal and the decrease in most of the other signals corresponding to the coformers [26].

Amides such as nicotinamide, isonicotinamide and hydroxybenzamides are often used as coformers of pharmaceutical multicomponent crystals [18, 27, 28]. However, it has been proved that, in the case of substituted analogues of benzamide-benzoic acid pairs, cocrystals are formed only if substitution of the benzoic acid moiety is done with electron-withdrawing functional groups [29]. This indicates that the ability of benzamide to form cocrystals is limited. Urea, on the other hand, can easily form molecular complexes in the solid state, as evidenced by the large number of cocrystal structures deposited in the Cambridge Structural Database (CSD) [30]. This high potential of cocrystal formation of urea can be explained by the fact that it possess both good hydrogen bond (HB) donor and acceptor capabilities. However, only a few cocrystal structures of urea with aromatic carboxylic acids can be found in CSD, for example with salicylic acid (SLCADC), 3,5-dinitrosalicylic acid (NUHYAQ), 1,1-binaphthyl-2,2′-dicarboxylic acid (ROGKOO), trimesic acid (CEKSIU), o-phthalic acid (NUHYIY) and 5-nitrosalicylic acid (NUHXUJ).

The aim of this paper was threefold. Firstly, a simple cocrystallization method developed by our group was applied to extend the cocrystallization landscape of urea and benzamide with carboxylic acids. Secondly, the applicability of post-quantum chemistry approach within COSMO-RS framework for theoretical cocrystal screening was considered. Finally, substituents effects on RCOOH···H$_2$NCOR' $R^2_2(8)$ heterosynthon stabilities and charge transfer were explained.

## Methods

### Chemicals

All chemicals in this study were of analytical grade and were used as received without further purification. Urea (U, CAS: 57-13-6), benzamide (B, CAS: 55-21-0), oxalic acid (OA, CAS: 144-62-7), benzoic acid (BA, CAS: 65-85-0), salicylic acid (SA, CAS: 69-72-7), 3-hydroxybenzoic acid (3HBA, CAS: 99-06-9), 2,5-dihydroxybenzoic acid (2,5DHBA, gentisic acid, CAS: 490-79-9), 2,6-dihydroxybenzoic acid (2,6DHBA, γ-resorcylic acid, CAS:303-07-1), acetylsalicylic acid (ASA, CAS: 50-78-2) and methanol (CAS: 67-56-1) were purchased from POCH (Gliwice, Poland). Malonic acid (MOA, CAS: 141-82-2), maleic acid (MEA, CAS: 110-16-7), fumaric acid (FA, CAS: 110-17-8), succinic acid (SUA, CAS: 110-15-6), glutaric acid (GA CAS: 110-94-1), 4-hydroxybenzoic acid (4HBA, CAS: 99-96-7), 3,5-dihydroxybenzoic acid (3,5DHBA, CAS: 99-10-5) and 2,4-dihydroxybenzoic acid (2,4DHBA, β-resorcylic acid, CAS:89-86-1) were obtained from Sigma-Aldrich (St. Louis, MO).

### Experimental procedure

The samples used for cocrystal screening were prepared via the DEC approach, as described in detail in our previous studies [24–26]. This simple procedure, validated and successfully applied for cocrystal screening, consists of the three following steps. Firstly, the methanolic solutions of the solid mixtures components were prepared and mixed together in equimolar proportions. Then, 20 μl of the mixtures and pure components solutions were placed on glass microscope slides and left to evaporate under conditions of 43 °C and atmospheric pressure. Finally, the crystallite layers thus formed were analyzed directly on the slide using PXRD and ATR-FTIR. These measurements were performed with the help of a Goniometer PW3050/60 equipped with an Empyrean XRD tube Cu LFF DK303072 and Bruker Alpha-PFT-IR spectrometer (Bruker, Karlsruhe, Germany) combined with an attenuated total reflection (ATR) diamond device.

### Computations

The computations performed relied on the assumption that intermolecular interactions responsible for formation of intermolecular molecular complexes in the solid state can be estimated reliably based on mixing thermodynamic functions characterizing the super cooled liquid solution obtained after mixing of components at a specific stoichiometric ratio under ambient conditions. Thus, negative values of mixing enthalpy or Gibbs free energy are thought to indicate that the mixture is thermodynamically favored over the pure component liquids. In other words, the miscibility of the supercooled liquid is thought to be also associated with miscibility in the solid state, hence showing the ability to cocrystallize. This hypothesis seems crude but was applied successfully to prediction of the probability of cocrystallization of several active pharmaceutical ingredients [31]. The excess thermodynamic functions characterizing hypothetical molecular compounds of a given stoichiometry requires estimation of the thermodynamic functions of coformers in their pure liquid state under supercooled conditions, and additional computations of their properties in a mixture composed of two components with given stoichiometric proportions. Hence, the mixing enthalpy can be defined as follows:

$$\Delta H_{12}^{mix} = H_{12} - \left(x_1 H_1^1 + x_2 H_2^2\right) \quad (1)$$

where subscript denotes solute and superscript represent solvent type. It is worth mentioning that the superscripts denote modeled rather than real solutions used during crystallization. Hence, the 1 or 2 superscripts denote single component liquids, while 12 represent the two-component mixture under supercooled conditions. The enthalpy of cocrystal formation, $H_{12}$, can be estimated based on computations in the bi-component liquid

$$H_{12} = x_1 H_{12}^1 + x_2 H_{12}^2 \quad (2)$$

The excess enthalpy accounts for all energetic contributions, including hydrogen bonding and van der Waals interactions of all energetically favorable coformers of each components. Thus, COSMOtherm [32] explicitly considers all possible contacts between these coformers, and provides excess thermodynamic functions after averaging over all the contributions. The Gibbs free energy associated with molecular complex formation can be related directly to excess enthalpy after considering entropic contributions:

$$\Delta G_{12}^{cc} = \Delta H_{12}^{mix} - T \Delta S_{12}^{mix} - \Delta \Delta G^{fus} \quad (3)$$

An additional assumption is that the difference between the free energy of fusion of the cocrystal and the reactants is insignificant, $\Delta\Delta G^{fus} \approx 0$. Hence, the Gibbs free energy can be obtained after estimation of the related chemical potentials at infinite dilution:

$$\Delta G_{12}^{mix} = G_{12} - \left(x_1 \mu_1^1 + x_2 \mu_2^2\right) + RT(x_1 \ln x_1 + x_2 \ln x_2) \quad (4)$$

$$G_{12} = x_1 \mu_{12}^1 + x_2 \mu_{12}^2 \quad (5)$$

Two levels of approximation arise from these equations, namely one including entropic contributions and an alternative one assuming that the entropy of mixing is also negligible. Thus, for further characterization of cocrystals and simple eutectics both these options were used and compared each to other. All results presented correspond to quantum chemical COSMO-RS calculations based on BP86 functional and def2-TZVPD basis set computations using Turbomole [33]. The sigma profiles were generated after single point calculations with fine gridding of the tetrahedron cavity, inclusion of a hydrogen bond interaction term (HB2012) and accounting for a van der Waals dispersion term based on the D3 Grimme method [34, 35]. Such an approach is considered the most advanced level available so far for predicting thermodynamic data using a combination of computational chemistry and statistical mechanics.

## Results and discussion

The cocrystallization of benzamide and urea with carboxylic acids can be represented by the model system of the RCOOH···H$_2$NCOR' R$_2^2$(8) heterosynthon. This supramolecular pattern occurs very frequently in solid materials and is found far more often in the CSD than the corresponding homosynthon. Since the experimental data are not very rich in systems of this type, the first part of this paper extends the collection of cocrystals by documenting the synthesis of new cocrystals. In the second part, a theoretical screening procedure utilizing an excess thermodynamics functions is used to characterize intermolecular interactions in the considered systems. Finally, the third part of the paper discusses quantification of substituent effects on synthon properties.

### Experimental exploration of the cocrystallization landscape

*Binary mixtures of urea with carboxylic acids*

Urea is quite often found as a former of cocrystals, and more than 100 solved structures of this compound with quite diverse coformers are known. This common occurrence of urea in cocrystals is related to its ability to form a variety of supramolecular patterns. The carbonyl group acts as a very strong acceptor center and can form mono- or bi-center hydrogen bonds. Additionally, two amine groups offer rich donor capabilities resulting either in direct interactions with coformers, or occluding it in the network of hydrogen patterns involving urea itself. This is also visible in the great variety of coformer ratio. Although a 1:1 ratio predominates in the majority of cocrystals, one can find proportions as high as 10:1 in the case of urea-2,12-tridecanedione (MISNOR). A list of all known cocrystals of urea with carboxylic acids is provided in Table 1. In our previous study [26], the cocrystallization landscape of urea was enhanced by several molecular complexes formed

Table 1 The experimentally verified[a] cocrystallization abilities of benzamide (B) and urea (U). The experimental characteristics of new cocrystals identified in this work are provided in supplementary material, as indicated

| Coformer | U[a] | source | B[a] | source |
| --- | --- | --- | --- | --- |
| Oxalic acid (OA) | + | UROXAM, UROXAL, Fig. S1 | + | This work, Fig. S7 |
| Malonic acid (MOA) | + | URMALN, Fig. S2 | + | This work, Fig. S8 |
| Maleic acid (MEA) | + | CEKRUF, CEKSAM, Fig. S3 | + | This work, Fig. S9 |
| Fumaric acid (FA) | + | TIPWIY, Fig. S4 | + | YOPBUB, Fig. S10 |
| Succinic acid (SUA) | + | UNIRT, VEJXAJ, Fig. S5 | + | BZASUC, Fig. S11 |
| Glutaric acid (GA) | + | ZODWIY, TONGOS, Fig. S6 | + | This work, Fig. S12 |
| Salicylic acid (SA) | + | SLCADC | + | URISAQ, Fig. S13 |
| 2,4-Dihydroxybenzoic acid (2,4DHBA) | + | [26] | + | This work, Fig. 1a,b |
| 2,5-Dihydroxybenzoic acid (2,5DHBA) | + | [26] | + | This work, Fig. S14 |
| 2,6-Dihydroxybenzoic acid (2,6DHBA) | + | [26] | + | This work, Fig. S15 |
| 3,5-Dihydroxybenzoic acid (3,5DHBA) | + | [26] | + | This work, Fig. S16 |
| Benzoic acid (BA) | − | [26, 36] | − | [29], Fig. S17 |
| Acetylsalicylic acid (ASA) | − | [26, 37] | − | This work, Fig. 1c,d |
| 3-Hydroxybenzoic acid (3HBA) | + | [26] | − | This work, Fig. S18 |
| 4-Hydroxybenzoic acid (4HBA) | + | [26, 38] | − | [29], Fig. S19 |

[a] Additionally, benzamide is known to be able to cocrystalize with pentafluorobenzoic acid (ESATUN), phenylacetic acid (MECHAF), 3-nitrobenzoic acid (OVEZUL), 4-nitrobenzoic acid (YOPCAI), 3,5-dinitrobenzoic acid (OVIBEB), and 4-hydroxy-3-nitrobenzoic acid (OVIBAX), while urea cocrystalizes also with 1,1′-binaphthyl-2,2′-dicarboxylic acid (ROGKOO), 2-((3-(3,4-dimethoxyphenyl)acryloyl)amino)benzoic acid (KINVAG), 2-Hydroxy-3,5-dinitrobenzoic acid (NUHYAQ), 2-phthalic acid (NUHYIY), 3-nitrobenzoic acid [39], 4-aminobenzoic acid (NUHYEU), 5-nitrosalicylic acid (NUHXUJ), adipic acid (ERIWUY), barbituric acid (EFOZAB), cis,cis-1,3,5-cyclohexanetricarboxylic acid (XORMUM), cyanuric acid (PANVUV), DL-tartaric acid (NEHPIZ), D-tartaric acid (NEZDAX), glycine (NUBHOH), heptanedioic acid (EVETAB), itaconic acid (PANVAB), parabanic acid (URPRBN10), pyrazine-2,3-dicarboxylic acid (NUHYOE), pyridine-2,6-dicarboxylic acid (NUHYUK) and suberic acid (QQBKM)

with aromatic carboxylic acids, as documented in Table 1. It is worth noting that dicarboxylic aliphatic acids, like oxalic, malonic, succinic and glutaric acid, are generally regarded as good coformers [40], so their ability to form molecular complexes with urea is not surprising. What is worth mentioning is that, occasionally and contrary to chemical intuition, urea cannot cocrystalize with some coformers. For example, it is not possible to cocrystalize urea with aspirin (acetylsalicylic acid)—a common prototypical analgesic having anti-inflammatory and antipyretic properties. However, the structurally very similar salicylic acid, which is also used as an analgesic, has been successfully cocrystalized with urea (SLCADC). The utilization of oriented samples confirmed cocrystallization of urea with carboxylic acids and also properly identified negative cases [26]. Examples of our measurements are provided in supporting materials, as indicated in Table 1.

*Binary mixtures of benzamide with carboxylic acids*

Benzamide structure differs from urea simply by replacement of one amine group with an aromatic ring. This, however, can seriously alter the possible hydrogen boding patterns due to three major factors. First, the lack of one donor side reduces the diversity of possible supramolecular patterns. Additionally, the bulky phenyl group introduces steric restrictions, making it more difficult to pack in multicomponent crystals. Finally, the resonance effect alters significantly electron densities on both carbonyl and amino groups. All this added together explains why benzamide is a much less common former of cocrystals. As documented in Table 1, there are only 17 known benzamide-carboxylic acid solids, including those proved by our experiments. The collected results of DEC are presented in the Supporting Material. However, exemplary positive and negative cases are discussed briefly here. Figure 1a shows the PXRD patterns and ATR-FTIR spectra obtained by direct on-glass measurements for a benzamide-2,4DHBA (B-2,4DHBA) cocrystal and benzamide-acetylsalicylic acid (B-ASA) mixture. In the former case, the registered signals characterizing the binary mixture confronted with patterns obtained for pure components unequivocally confirm new compound formation by identification of new peaks corresponding to a cocrystal phase.

Interestingly, the miscibility of benzamide and 2,4DHBA in the solid phase was also confirmed by ATR-FTIR spectroscopy, since the consequence of formation of the carboxylic acid amide synthon is a visible shift of N–H vibration in comparison to pure amide spectra [41]. In the case of B-2,4DHBA, the absorption band corresponding to stretching of the N−H vibration mode was shifted from 3366 cm$^{-1}$ to 3417 cm$^{-1}$.

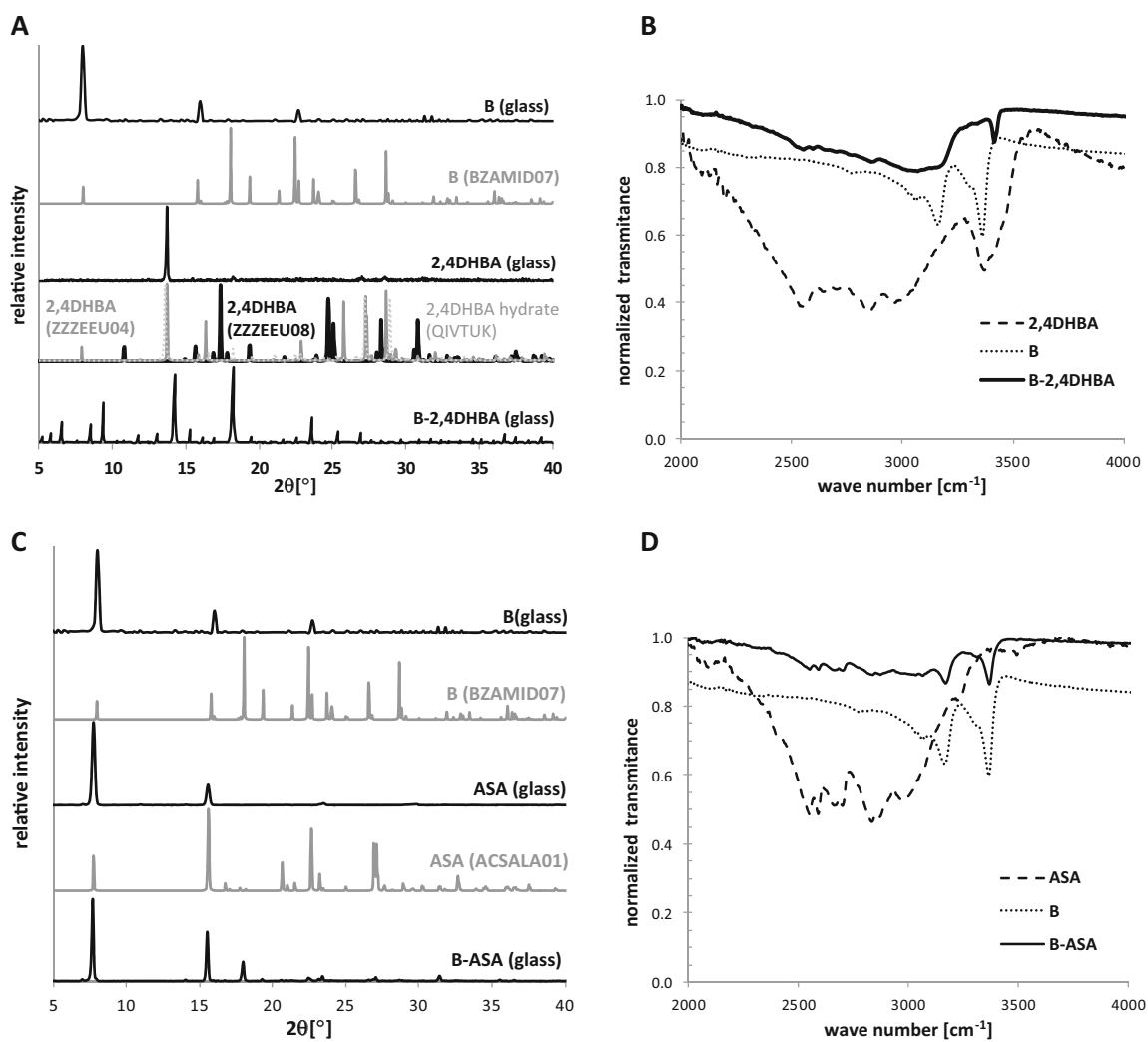

**Fig. 1** Powder X-ray diffraction (PXRD) patterns and attenuated total reflectance Fourier transform infrared spectroscopy (ATR-FTIR) spectra of benzamide-2,4DHBA (B-2,4DHBA) cocrystal (**a**, **b**) and benzamide-acetylsalicylic acid (B-ASA) mixture (**c**, **d**) obtained via droplet evaporative crystallization (DEC) on a glass surface. Spectra of pure components are also shown

Such shifts of absorption bands can be observed also in other cases where a new cocrystal phase is formed, which can be inferred directly from figures provided in the Supporting Material (Figs. S1–S19). The second example provided in Fig. 1 clarifies that the lack of crystal structure in the CSD database for the B-ASA pair is not accidental. As can be seen from Fig. 1c, in the case of the benzamide-acetylsalicylic acid system, an overlapping of single component PXRD patterns with signals measured for mixtures of benzamide with ASA is evident. Furthermore, there are no absorption bands shifts in this case, which additionally indicates the probable immiscibility of components in the solid state (Fig. 1d). Similar observations can lead to the conclusion that mixtures of benzamide with 3HBA, 4HBA and BA also show immiscibility in the solid state (see Figs. S17–S19). It is worth mentioning that the lack of cocrystal formation in the latter two cases has already been documented [29].

## Theoretical cocrystal screening

The statistical thermodynamics approach offered by post-quantum analysis has the advantage of providing values of thermodynamic functions. From the perspective of this project, the excess enthalpies and excess Gibbs free energies are of particular importance. The observed diversity of pair behaviors requires an unambiguous definition of the criterion for selecting excess data for further analysis. One can take just the values corresponding to 1:1 molar proportions of components, ignoring the fact that cocrystals can be formed with different stoichiometry. Alternatively, one can take into account the shape of $H^{mix}(x_1)$ or $G^{mix}(x_1)$ and locate molar fractions corresponding to the minima found on such curves. Figure 2 presents some exemplary plots. Although the concentration-dependent mixing enthalpies and Gibbs free energies are quite similar for oxalic acid, the ones characterizing benzoic acid, 2,6-dihydroxybenzoic acid, 4-aminobutanoic acid and many other carboxylic acids are quite diverse. They can differ in shape, sign and localization of the extrema points. Such incongruent trends of $H^{mix}$ and $G^{mix}$ suggest that the entropic contributions to thermodynamics of mixing are often non-negligible. The set of binary mixtures used for virtual cocrystal screening of benzamide or urea comprised 514 aromatic carboxylic acids found in CSD as constituents of bi-component cocrystals. The complete list of analyzed pairs, along with the values of $H^{mix}$ and $G^{mix}$ obtained are provided in the Supporting Material (Table S1). In Fig. 3 these values are presented graphically by providing the corresponding distribution plots. The first observation is that, for these types of systems, the curves characterizing values of Gibbs free energy are shifted more toward negative values than those derived using the values of excess enthalpies. On the other hand, on smoothed histograms classifying the distributions of excess enthalpies, the maximum is found in fairly the same place for both amides interacting with carboxylic acids. Besides, slightly more homogeneous interactions are to be expected in the case of benzamide since broader peaks associated with excess function of urea suggest a higher diversity of interactions in such case.

Consequently, one can find pairs showing quite strong attractions along with the pairs for which repulsion predominates. Figure 4 presents the cumulative distributions of both $H^{mix}$ and $G^{mix}$ values, characterizing all 514 pairs of urea with carboxylic acids. Additionally, the values of excess thermodynamic functions for experimentally studied systems are provided. The conclusions drawn from Gibbs free energy distributions and from enthalpy of mixing are quite diverse. First of all, the percentage of structures exhibiting attractions is much higher when $G^{mix}$ is used as a quantification criterion, reaching 95 % versus 53 % for $H^{mix}$. Also, down-shifting of all quartiles by about 0.5 kcal mol$^{-1}$ if $G^{mix}$ is used is also worth noting. Also the medians differ by the same interval. The straight lines presented in Fig. 4 denote the excess values for all experimentally studied pairs. From the perspective of $H^{mix}$, the two known simple eutectics of urea are found within the first two quartiles, and all known cocrystals are outside of this range, belonging to the third or fourth quartiles. Similar

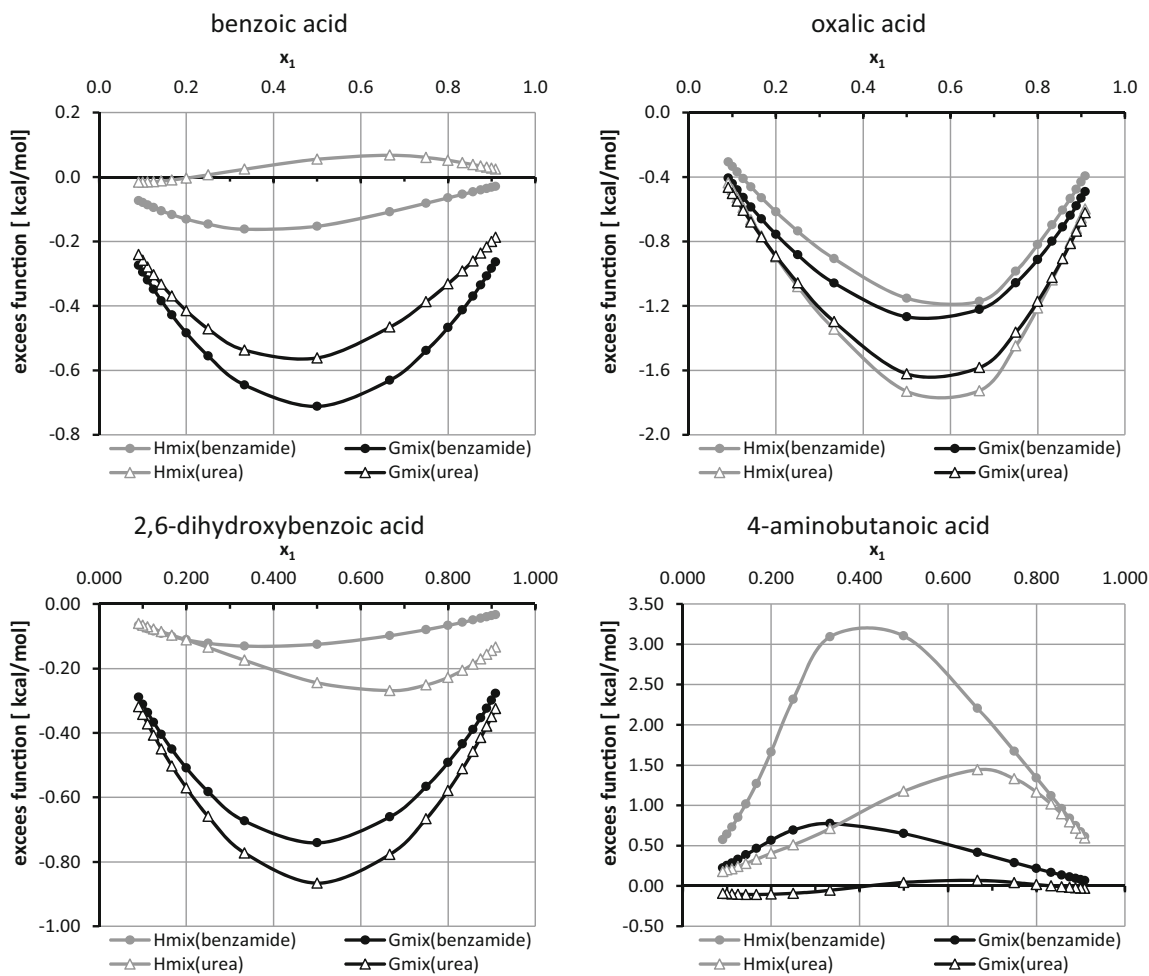

**Fig. 2** Examples of typical plots of $H^{mix}(x_1)$ or $G^{mix}(x_1)$ corresponding to supercooled fluids of coformers involved in cocrystallization of urea or benzamide with carboxylic acids

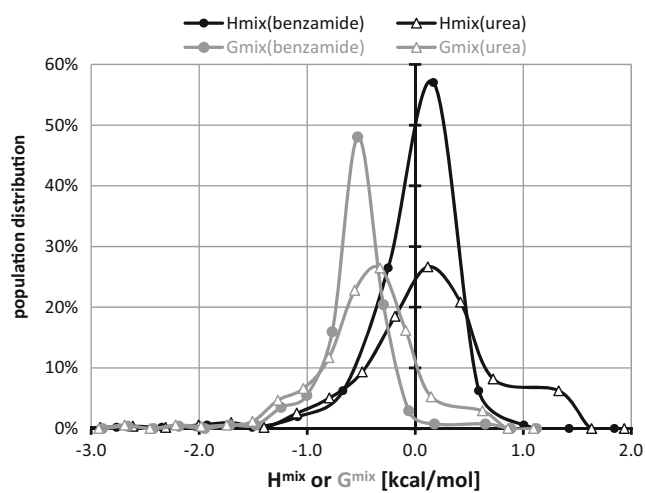

**Fig. 3** Smoothed histograms characterizing distributions of $H^{mix}$ and $G^{mix}$ of 514 pairs of carboxylic acids interacting with benzamide or urea

conclusions can also be drawn using $G^{mix}$. This suggests that classifying a potential pair as falling into a cocrystal group or simple eutectic group is fairly reliable in the case of urea interacting with carboxylic acids due to its quite distinct energetic patterns. Turning our attention to benzamide pairs and inspecting the data presented in Fig. 5, no such uniquivocal conclusion can be drawn. First of all, the percentages of structures exhibiting attraction in terms of negative values of $G^{mix}$ and $H^{mix}$ are much closer to each other, at 99 % and 82 %, respectively. The higher percentage of structures with negative values of excess thermodynamic functions is also reflected in the changes in the quartiles and median values. This makes it much more complicated for the separation of samples showing miscibility in the solid phase from those not being able to form homogenous solid dispersions. Indeed, the simple eutectics formed by benzamide are characterized by $H^{mix}$ and $G^{mix}$ almost within the same range as real cocrystals. This somewhat negative conclusion limits, to some extent, the generality of cocrystal screening based on liquid mixtures under supercooled conditions. As already mentioned [42], based on distribution of heat of formation in water, where a significantly lower percentage of structures were confirmed after applying the proposed scoring function for this particular group of compounds, the prediction of aromatic amides cocrystals is much more difficult than that of carboxylic acids.

### Substituent effects on –COOH⋯H$_2$NOC- hetero synthon properties

The major energetic contribution to stabilization of benzamide and urea cocrystals with carboxylic acids comes from –COOH⋯H$_2$NOC- heterosynthons of the $R^2_2(8)$ type. The sensitivity of this pattern to substituent effects can be revealed directly in the case of benzoic acid analogues. This can be done by studying relationships with respect to the Hammett constant values [43] characterizing the electro-donating and electro-withdrawing capabilities of substituents bonded at the para position. Unfortunately, there is no detectable trend for computed excess thermodynamic functions with Hammett constants, but application of an alternative strategy can be quite efficient. For this purpose, full gradient optimization was performed for 110 pairs of para-substituted benzoic acid analogues with benzamide and urea. The list of these species is provided in the Supporting Material as Table S2. The geometries of all of these pairs were fully optimized by means of the ωB97XD density functional with 311++G** basis set, as implemented in the GAUSSIAN09 package [44]. In order to characterize the contribution to pair stabilization energy, the decomposition analysis was performed based on the absolutely localized orbitals approach (ALMO) [33] implemented in the QChem 4 package [45, 46]. This method offers decomposition of the total intermolecular binding energy into several components including, among others, a charge-transfer (CT) portion of binding energy [45]. Because formation of hydrogen bonds in the intermolecular complexes is often associated with significant charge transfer between monomers, the ALMO method can be used to provide a conceptual description of an amount of bidirectional charge transfer from and to interacting monomers. The results of such decomposition of the total charge-transfer into complementary occupied-virtual orbitals is typically expressed as the sum of four contributions to the total electron charge dislocation coming from the intra and inter CT of both components [47]. The results of binding energy $\Delta E^{BSSE}$ computations and charge transfer analysis are provided in Figs. 6 and 7, respectively. It is interesting to note that the observed substituent effects on binding energy are quite strong and highly linear. The greater the electron-withdrawing character of the substituent, the stronger the binding of aromatic carboxylic acids with either of the considered amides. This trend is slightly more pronounced in the case of urea, suggesting not only a higher sensitivity to substituent character but usually also stronger stabilization of the considered heterosynthon. It is interesting to note that interactions between the carboxylic and amide groups involve two types of hydrogen bond formation, with –COOH playing both a donor and acceptor role. As the consequence of such contacts, charge transfer is allowed from and to urea or benzamide via these two hydrogen bonding channels. According to expectation, the greater the electron-withdrawing character of the substituent, the stronger the CT from amide to benzoic acid analogue. On the other hand, the opposite direction of CT is also possible but, as documented in Fig. 7, it is much smaller. Indeed, the partial charge transferred from aromatic carboxylic acid towards the amide stands at roughly half of the most dominant contribution observed for the reverse direction. All trends mentioned are highly linear for both amides paired with para-substituted benzoic acid analogues. It seems that complexes involving urea are more polarizable compared to those comprising benzamide.

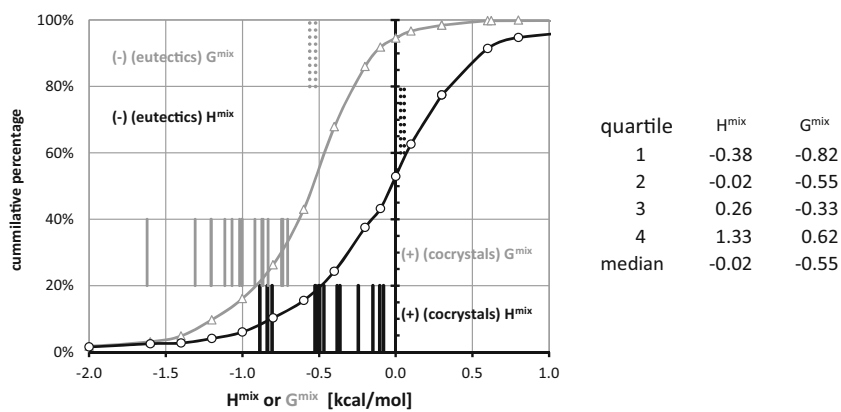

**Fig. 4** Cumulative distributions of urea-aromatic carboxylic acid pair populations expressed as a function of increasing inclusion criterion $H^{mix}$ (*black lines*) or $G^{mix}$ (*gray lines*). The *lines* represent energetics of mixing of urea with carboxylic acids for which the ability of cocrystallization or simple binary eutectics formation is experimentally validated

| quartile | $H^{mix}$ | $G^{mix}$ |
|---|---|---|
| 1 | -0.38 | -0.82 |
| 2 | -0.02 | -0.55 |
| 3 | 0.26 | -0.33 |
| 4 | 1.33 | 0.62 |
| median | -0.02 | -0.55 |

**Fig. 5** Cumulative distributions of populations of benzamide-aromatic carboxylic pairs expressed as a function of increasing values of $H^{mix}$ (*black lines*) or $G^{mix}$ (*gray lines*). The *lines* represent energetics of mixing of benzamine with carboxylic acids for which the ability of cocrystallization or simple binary eutectics formation is experimentally validated

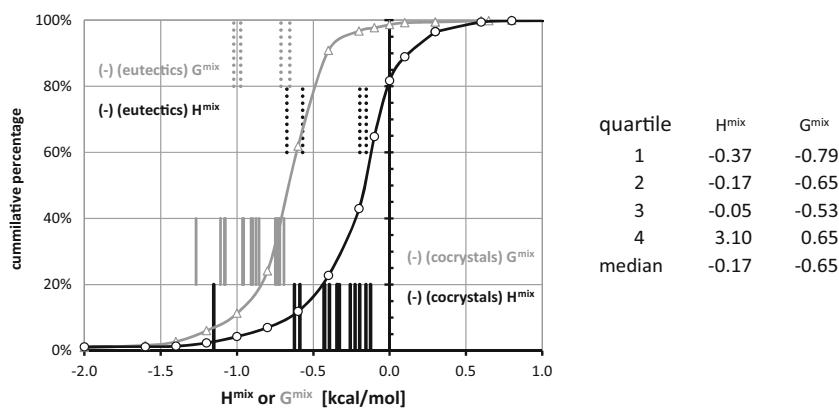

| quartile | $H^{mix}$ | $G^{mix}$ |
|---|---|---|
| 1 | -0.37 | -0.79 |
| 2 | -0.17 | -0.65 |
| 3 | -0.05 | -0.53 |
| 4 | 3.10 | 0.65 |
| median | -0.17 | -0.65 |

## Conclusions

A knowledge of cocrystallization abilities and the possibilities of simple binary eutectics formation are important issues having many practical implications. Here, two representative cases of popular coformers were analyzed in detail from both experimental and theoretical perspectives. Despite the significant structural similarities of the two considered amides, the consequences of mixing with carboxylic acids are quite diverse. The lack of one amino group in benzamide compared

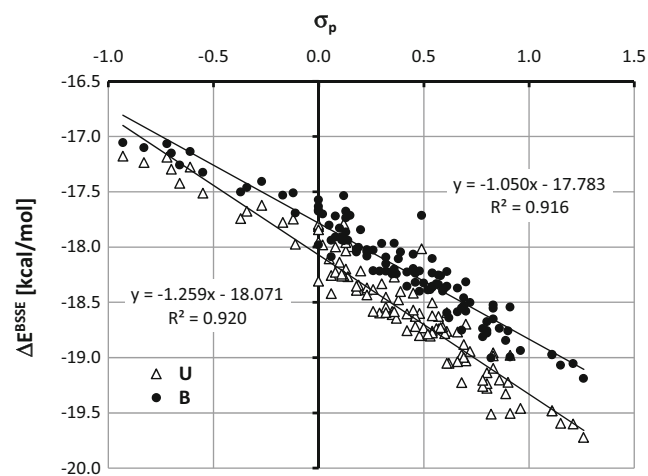

**Fig. 6** Substituent effects on binding energy of the $R^2_2(8)$ heterosynthon formed between urea or benzamide with para-substituted benzoic acid analogues ($\sigma_p$ stands for Hammett constant [43])

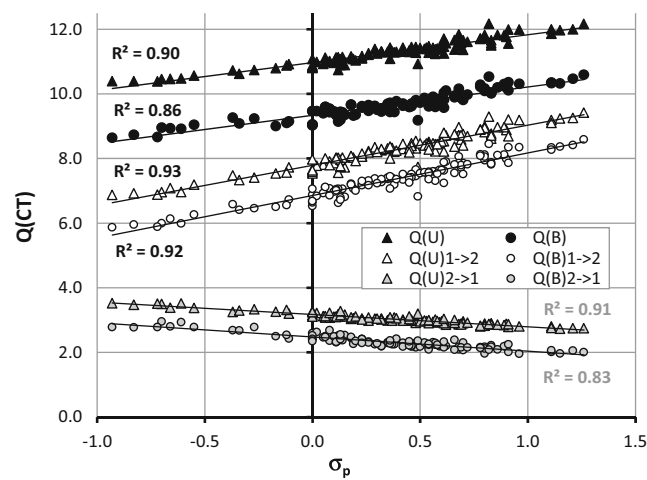

**Fig. 7** Substituent effects on a charge transfer between urea or benzamide and para substituted analogues of benzoic acid. The amount of charge transfer (CT) from amide to aromatic carboxylic acid and vice versa is represented by $Q_{1\rightarrow 2}$, $Q_{2\rightarrow 1}$; $\sigma_p$ stands for Hammett constant [43]

to urea results in a significant alteration of the intermolecular interactions with carboxylic acids despite potentially using the same synthon pattern. The profiles of smoothed histograms suggested that slightly more potential pairs of benzamide are characterized in the attractive region compared to urea. Thus, one could expect to find more cocrystals of benzamide than of urea. The experimental data are in opposition to this expectation, which is obvious if one compares 17 cocrystals of benzamide with 32 cocrystals of urea with carboxylic acids. However, the number of coformers known from literature data and this work is quite limited and hence further experimental effort is required. Nevertheless, it seems that discrimination of simple binary eutectics from the two component cocrystal is much more difficult in the case of benzamide than for urea interacting with carboxylic acids. One reason for this is the much less populated eutectic set for urea, which encompasses only two cases. The origin of this discrepancy might be related to the much higher divisibility of the hydrogen bonding pattern that can be potentially adopted by urea compared to that of benzamide. This emphasizes the necessity of individual characterization of the cocrystallization landscape for each particular compound rather than formulating general rules governing the affinities of two interacting coformers.

**Acknowledgments** This work utilized the COSMOtherm software with the BP-TZVPD-FINE database kindly provided by COSMOlogic. The project was partly supported by PL-Grid Infrastructure. The allocation of computational facilities of the Academic Computer Centre "Cyfronet" AGH/Krakow/POLAND is also acknowledged.